\documentstyle [aps]{revtex}
\begin{document}

\title{Decoherence~of~Quantum~Fields: Pointer~States~and~Predictability}

\preprint{LA--UR 95--3364}

\author{J.R.~Anglin and W.H.~Zurek}

\address{Theoretical~Astrophysics, T-6, Mail~Stop~B288,
Los~Alamos~National~Laboratory, Los~Alamos, NM~87545}

\date{\today}

\maketitle

\abstract{We study environmentally induced decoherence of an
electromagnetic field in a homogeneous, linear, dielectric medium.  We
derive an independent oscillator model for such an environment, which is
sufficiently realistic to encompass essentially all of linear physical
optics.  Applying the ``predictability sieve'' to the quantum field,
and introducing the concept of a ``quantum halo'',  we recover the
familiar dichotomy between background field configurations and
photon excitations around them.  We are then able to explain why a
typical linear environment for the electromagnetic field will
effectively render the former classically distinct, but leave the
latter fully quantum mechanical.  Finally, we suggest how and why
quantum matter fields should suffer a very different form of
decoherence.}

\draft

\section{Introduction}

Decoherence and environmentally induced superselection have been
studied extensively in the system composed of a single harmonic
oscillator linearly coupled to a bath of independent
oscillators\cite{GSI,CaldeiraLeggett,HPZ}.  This system has generally
been presented as a conveniently solvable model of value in
investigating fundamental problems of principle, such as the issues of
dissipation in quantum mechanics\cite{Schwinger,Ullersma}, or of
emergence of classical behaviour in open systems\cite{UnruhZurek,ZHP}.
In this paper we point out that this simple system actually
constitutes a realistic description of a quantum electromagnetic field
propagating in a linear dielectric medium.  The mechanisms of
decoherence identified in single oscillator models can therefore be
applied straightforwardly to electrodynamics.

The particular aspect of decoherence that we consider is the selection,
by the environment and its coupling to the system, of a preferred basis
of {\it pointer states}\cite{Zurek1}.  We find that the linear
interaction of the electromagnetic field with the environment implies
that the pointer states of the quantum field are coherent states.
While single oscillator models often tend to suggest the interpretation
of coherent states as localized particles, in the case of the field
they are not localized photon packets at all: the pointer states of the
quantum electromagnetic field are in fact background field
configurations.

There are also, however, many experiments which reveal the existence of
photons; and so we examine decoherence in our model more carefully, to
determine how it is that photons can be robust despite propagating
through an environment.  We are led to associate with every pointer
state a {\it quantum halo} of states that are not effectively
distinguished from it by the environment, and we show that excitations
of a few photons above a background field are typically states within
such a quantum halo.

The paper is organized as follows.  The following section presents our
model system, and derives a description of a typical dielectric
medium as a bath of independent oscillators, from the assumption that
such a medium will contain a large number of molecules within a
volume on the scale of the smallest electromagnetic wavelength
under study.  We then specialize considerably to the case of
ultraweak Ohmic dissipation at ultrahigh temperatures.  In
Section II, we take advantage of this simplification to derive
several exact results concerning the pointer states of our system.
Our third section then discusses quantum halos.  Section IV then
summarizes our results, and briefly suggests why decoherence may be
expected to affect matter fields much differently from linearly coupled
systems such as the electromagnetic field.

\section{The Model}

The system we will study will be an electromagnetic field in 3+1
dimensions.  We quantize the field in Coulomb gauge, in a box of linear
dimension $2L$,
and couple it to molecules composing a dielectric medium inside the box:
\begin{equation}\label{Ls}
{\cal L}_{\cal S} = {1\over2}\sum_{s=1}^2\sum_{\vec k=-\vec K}^{\vec K}
\Bigl[\dot{A}_{\vec{k},s}^2 - \omega(k^2)
A_{\vec{k},s}^2
+ g\sum_n \dot{A}_{\vec{k},s}
e^{-i{\pi\over L}\vec{k}\cdot\vec{x}_n} j_{n,s}\Bigr]
\;.
\end{equation}
We let $\vec{k}$ label the Fourier modes and $s$ the polarization
states; $\vec{K} \equiv (L\Gamma,L\Gamma,L\Gamma)$, where $\Gamma$ is
an ultraviolet cut-off wave number, above which we consider the gauge
field to decouple from the medium (or at least to interact with it in
such a way that there will be negligible effect on the field modes
below the cut-off).  The quantities $j_{n,s}$ represent the electric
dipole interaction with molecules located at positions $\vec{x}_n$; $g$
is the coupling strength of this interaction, which is assumed to be
the same for all $n$ and to be small.

Turning now to the environment, we will initially assume merely that it
consists of a large number of molecules, which interact with each other
only via the gauge field coupling presented above, and which are
located at the points $\vec{x}_n$.  We will neglect the motions of the
molecules (with consequences that may be easily remedied, as discussed
below), and consider only their internal energies:
\begin{equation}\label{hame}
\hat{H}_{\cal E} = \sum_{n} \hat{H}_n\;,
\end{equation}
where $\hat{H}_n$ have some arbitrary discrete spectra.  We will {\it
not} assume that the environment actually consists of independent
harmonic oscillators, but instead we will derive the fact that a
generic environment may be treated as such, in the limit of large
$N$\cite{FeynmanVernon}.
The $N$ that must be large is the number of molecules within a volume
on the cut-off scale; we will therefore require that the number density
of the medium satisfy $d>>\Gamma^3$.  For an ideal gas at room
temperature and atmospheric pressure, $d\simeq 10^3\Gamma^3$
corresponds to a cut-off of electromagnetic modes in the high
ultraviolet range ($\lambda \sim 10$ nm).  In solids or liquids we
might perhaps handle somewhat shorter wavelengths, but our derivation
of the independent oscillator model as a large $N$ approximation to a
general non-conducting environment must be expected to break down in
the X-ray band ($\lambda \sim 1$ nm).

We will treat the medium as an unobserved environment, and describe
only the state of the electromagnetic field, using the reduced density
matrix formed by tracing over the states of the environmental
molecules.  If we assume that the initial state is a direct product of
field and medium states, then we can obtain the evolution of the
reduced density matrix from the path integral propagator
\begin{equation}\label{rdmprop}
\rho[A,A';t] = \int\!{\cal D}A{\cal D}A'\,
\rho[A,A';0]\,e^{{i\over\hbar}\Bigl(S[A]-S[A']\Bigr)}\,F[A,A';t]\;,
\end{equation}
where $F[A,A']$ is the influence functional\cite{FeynmanVernon}
describing the effect of environmental molecules on the electromagnetic
field.  With the Hamiltonian (\ref{hame}), and a thermal initial state
for all the molecules, the influence functional is given by
\begin{eqnarray}\label{IFdef}
F[A,A';t] &=& {\rm Tr}_{\cal E}\Bigl(T
\exp\bigl[-{i\over\hbar}\sum_{n,\vec{k},s}\int_0^t\!dt'\,
\dot{A}_{\vec{k},s}(t')
e^{-i{\pi\over L}\vec{k}\cdot\vec{x}_n}\hat{\jmath}_{n,s}(t')\bigr]
		\nonumber\\
& &\qquad\times\;
\exp\bigl[-\sum_{n}\beta_n\hat H_n}\;
\bar Te^{{i\over\hbar}\sum_{n,\vec{k},s}\int_0^t\!dt'\,
\dot{A}'_{\vec{k},s}(t')
e^{-i{\pi\over L}\vec{k}\cdot\vec{x}_n}\hat{\jmath}_{n,s}(t')\bigr]
\Bigr)\;,
\end{eqnarray}
where $A_{\vec{k},s}(t)$ and $A'_{\vec{k},s}(t)$ are c-numbers in the
path integral for the field, but $\hat{\jmath}_{n,s}(t)$ is the dipole
moment operator of the $n$th molecule, in the interaction picture.  $T$
and $\bar T$ denote time-ordering and anti-time-ordering, respectively,
while $\beta_n$ is the usual inverse temperature, which we allow to
vary from place to place in the environment.  The trace is to be taken
over the states of the environment only.

We can now reduce this very general influence functional to the special
form of an independent oscillator model, by implementing our large $N$
approximation.  We divide the box of volume $8L^3$ into cells of volume
$\alpha^3\Gamma^{-3}$, where $\alpha$ is a number much smaller than
one.  Within the cell $C$ centred at the point $\vec{x}_c$ there will
be a large number $N(\vec{x}_c) = d(\vec{x}_c)\alpha^3\Gamma^{-3}$ of
molecules.\footnote{
	The appearance of $\alpha$ here would seem to lower, perhaps by
	an order of magnitude, the maximum frequencies up to which our
	analysis will be accurate.  As we will discuss below, however,
	it is easy to dispense with $\alpha$, which is only
	present to ensure that $e^{i\vec{k}\cdot\vec{x}}$ varies
	negligibly within a cell.}
By using time-dependent perturbation theory in the interaction picture,
keeping explicitly only terms up to second order in $g$, and zeroth
order in $\alpha$, we can obtain a simple form for the influence
functional for a single cell of dielectric medium:
\begin{eqnarray}\label{approx}
F[A,A';t] &=& \prod_{\vec{x}_{n}\in C}\Bigl(
	1 - {g^2\over\hbar^2}\bigl(\sum_j e^{-\beta E_j}\bigr)^{-1}
	\sum_{l,l'}\vert J_{ll'}\vert^2\nonumber\\
&&\qquad\times\sum_{\vec{k},s}
	\int_0^t\!dt'\int_0^2\!dt''\,e^{-\beta_c E_l}
	[\dot{A}_{\vec{k},s}(t') - \dot{A}_{\vec{k},s}(t')]\nonumber\\
& &\qquad\qquad\times [\dot{A}_{\vec{k},s}^*(t'')
	e^{-i\omega_{lm}(t'-t'')} - \dot{A}_{\vec{k},s}'^*(t'')
	e^{i\omega_{lm}(t'-t'')}]\Bigr)\;.
\end{eqnarray}
Here $J_{ll'}$ are unpolarized matrix elements of the dipole moment
operator, {\it i.e.}, we assume unpolarized scattering from individual
molecules, so that at the initial time $t=0$
\begin{equation}
\langle E_l\vert\hat{\jmath}_{n,s}\hat{\jmath}_{n,s'}\vert E_{m}\rangle
= \delta_{s,s'} \sum_{m} J_{lm}J_{ml'}
\;.
\end{equation}
There are no terms linear in $g$, because we take our molecules to have
no preferred orientation of their dipole moments:
$\langle\hat{\jmath}_{n,s}\rangle = 0$.  And we assume that the initial
state of the environment is a direct product of single-molecule thermal
states, with every molecule in a cell having the same initial
temperature $(k_B\beta_c)^{-1}$.

{}From the last line of (\ref{approx}) we discard all but the leading
terms in $N(\vec{x}_c)$, then put all the cells together and smooth out
the cell structure by defining interpolated density and inverse
temperature fields $d(\vec{x}), \beta(\vec{x})$.  We can even allow the
molecular composition of the environment to vary from cell to cell, so
that the entire form of $I_{eff}$ is spatially dependent as well.  We
find that the influence functional for the dielectric medium is that of
a set of independent harmonic oscillators at every point in the box,
(which we can now allow to become infinite):
\begin{eqnarray}\label{optics}
F[A,A';t] &=& \exp\Bigl( -{g^2\over2\hbar}\sum_s \int\!d^3x\,d(x)
   \int_0^\infty\!{d\omega\over\omega}\,I\bigl(\beta(x),\omega;x\bigr)
		\nonumber\\
&&\qquad\times\int_0^t\!dt'\int_0^{t'}\!dt''\,
	[\dot{A} - \dot{A}]_{t'}\bigl([\dot{A} - \dot{A}]_{t''}
	\coth{\hbar\beta(x)\omega\over2}\cos\omega (t'-t'')\nonumber\\
&&\qquad\qquad\qquad\qquad\qquad
-i[\dot{A} + \dot{A}]_{t''}\sin\omega (t'-t'')\bigr)\Bigr)\;,
\end{eqnarray}
where the spectral density of the effective bath of independent
oscillators is the (generally) temperature dependent quantity
\begin{equation}\label{Ieff}
I\bigl(\beta(x),\omega;x\bigr) =
	{4\omega\sinh{\hbar\beta(x)\omega\over2}
\sum_{l,m}\vert J_{lm}(x)\vert^2\;
\delta\bigl(\omega- {E_l(x)-E_m(x)\over\hbar}\bigr)
\over\hbar\sum_{l}e^{-\beta(x) E_l(x)}}\;.
\end{equation}

This effective environmental model describes physical optics in linear
dielectric media, at all frequencies below the cut-off, and for all
field strengths below thresholds for current generation.  The failure
of our model to describe conductors and non-linear media is clearly due
to our neglect of charge motion and higher-order terms in ${1\over N}$,
and so our recovery of linear optics is not simply a co-incidence.  In
the important and prevalent cases where free charges and non-linear
effects are negligible, our result is indeed physically sound, even
though our derivation may have appeared somewhat naive.  In particular,
our assignment of fixed positions to the molecules is certainly a very
crude treatment, especially for gases; but our results can be checked
by comparison with a more sophisticated analysis, in which the
molecules are treated as an ideal gas whose initial state is described
by a grand canonical ensemble.  The only additional effects one finds
are thermal broadening of the molecular spectra, and a Gaussian cut-off
on the effective coupling of field modes with energies on the scale of
the temperature (reflecting the smaller number of molecules possessing
kinetic energies in this range).

This more sophisticated analysis must assume that the gas is dilute, so
that quantum statistics are not significant, as well as that
$d\Gamma^{-3} >> 1$.  It is worth noting that the fuller analysis does
not require $d\alpha^3\Gamma^{-3} >>1$ for some small $\alpha$: the
delocalization of the molecules will itself smear out the phases
$e^{i\vec{k}\cdot\vec{x}_n}$, so that the smallest volume containing
very many molecules need only be on the cut-off scale, and not so much
smaller still that $e^{i\vec{k}\cdot\vec{x}_n}$ varies negligibly
across it.  This effect of the fuller treatment can be incorporated in
an approach like ours above, by making the $x_n$ into stochastic
variables, which fluctuate over distances on the order of
$\Gamma^{-1}$.  In the influence functional, we can then take the
ensemble averages of all the locations, and obtain Eqn. (\ref{optics})
even when $\alpha\to 1$.  It is thus evident that the delocalization
of molecules that obviates $\alpha$ need not be coherent.  For solids
and liquids delocalization is not so obviously sufficient to eliminate
$\alpha$, but since they are denser, we can retain $\alpha$ and still
achieve a cut-off in the high UV range.

Thermal broadening and cut-offs can also be incorporated
phenomenologically, and so we have presented the cruder analysis with
fixed molecular positions, in order to more clearly make the physical
point that large numbers of molecules within a cut-off volume leads to
effectively linear behaviour of an environment.  (It is also in aid of
this demonstration that we have been careful to employ the infra-red
regulator $L$, for if we had assumed from the start a countable number
of molecules and a continuum of field modes below any UV cut-off, we
could never have achieved the correct high ratio of molecules to
modes.  In this instance, the IR regulator is not just mathematical
pedantry, but is actually necessary to express some important physics.)

Since thermal motion and various sources of dissipation on the
molecular excitations will broaden the spectral lines, we will assume
that $I(\omega,\beta;x)$ is a continuous function of $\omega$ ---
though it may have sharp peaks around strong absorption lines.  This
will have the unphysical effect of giving the environment an infinite
specific heat capacity, so that radiative heating and cooling will be
neglected; but for most optical phenomena, and for the subjects
discussed in this paper, this will not be important.

The model we have arrived at encompasses all the physics of reflection
and refraction, and absorption.  It provides
\begin{equation}\label{ImK}
{\rm Im}K(\omega,\beta;x) = {\pi g^2\over2\Omega} I(\omega,\beta;x)\;,
\end{equation}
where $n(\omega,\beta;x)\equiv\sqrt{K(\omega,\beta;x)}$ is the complex
index of refraction.  The real part of $K$ is given, as it should be
for a linear medium, by the Kramers-Kronig relation
\begin{equation}\label{K-K}
{\rm}K(\omega,\beta;x) = 1 + {2\over\pi}\int_0^\infty\!d\omega'\,
{\omega'\over \omega'^2-\omega^2}\,{\rm Im}K(\omega',\beta;x)\;,
\end{equation}
taking the Cauchy principal part of the integral.  (The formal
derivation of these results is straightforward; the relation between
the quantum theory and classical optics will be clarified in the
remainder of this paper.)  Our model also describes thermal radiation,
albeit without heating or cooling of sources and sinks.

Nevertheless, for simplicity in the remainder of this paper we will
assume perfect spatial homogeneity.  In this limit, the Fourier modes
of the field decouple, even though they interact with the environment.
Each field mode thus constitutes a harmonic oscillator linearly coupled
to its own private bath of independent oscillators, with a continuous
spectral density.  And so we obtain a conclusion which will allow us to
apply the results of many apparently idealized studies of decoherence
to a real and important physical phenomenon: electrodynamics in a
homogeneous linear dielectric medium is, within the physically tenable
assumptions and approximations we have made, a realization of harmonic
quantum Brownian motion in the independent oscillator model.

\section{Pointer states}

Having mapped our field theoretic problem onto the problem of harmonic
Brownian motion in an independent oscillator environment, we are now
able to determine the pointer states of the field, in a straightforward
way.  We first review a clear-cut procedure for identifying pointer
states:  the {\it predictability sieve}\cite{Zurek2,ZHP}.  We extend
slightly the argument of Ref. \cite{ZHP}, in which certain squeezed
states are shown to minimize linear entropy, and also to yield the
smallest von Neumann entropy generation among all Gaussian initial
states.  Here we show that these same states actually minimize von
Neumann entropy against unrestricted variations of the initial states.

Pointer states are those states which are preferred by decoherence, in
a process that may be termed ``environmentally induced
superselection''.  A generic quantum state will tend to evolve into a
probabilistic mixture of pointer states.  The suppression of quantum
interference between these states makes the parameter space of pointer
states the natural phase space of the classical limit of the quantum
system in question.  The predictability sieve identifies the pointer
states by demanding that the environmentally induced splitting of a
quantum state into non-interfering branches be stable: the branches
must not rapidly branch in their turn.  A pointer state must remain as
pure as possible despite environmental decoherence.

A concrete expression of this requirement is that pointer states
minimize the growth of the entropy.  We therefore wish to use our
propagator (\ref{rdmprop}) to compute the reduced density operator
$\hat{\rho}(t)$ that evolves from some pure initial state with wave
function $\psi_i$.  From this we will obtain the von Neumann entropy
$S(t) = - \rm{Tr} \hat{\rho}(t)\ln\hat{\rho}(t)$ of this density
operator, as a functional of $\psi_i$.  Extremizing $S(t)$ with respect
to variations of $\psi_i$ then identifies those initial states that
acquire the least entropy by time $t$.  Since we must ensure that our
variations maintain the normalization of the initial state, we must
solve the constrained variational problem
\begin{equation}\label{Smin}
\rm{Tr}\Bigl[ (\ln\hat{\rho} + 1){\delta\hat{\rho}\over\delta\psi_i} =
\lambda \psi_i^*\Bigr]\;,
\end{equation}
for some Lagrange multiplier $\lambda$.

In general, entropy evolves in a complicated way during Brownian
motion, and this procedure becomes too difficult; but since we are
concerned here with decoherence, and not with such other effects as
dissipation and thermalization, we select the simple model which has
Ohmic spectral density and in which the dissipation rate $\gamma\to
0$.  We let the temperature become infinite, such that $\gamma T$
remains finite, and the environmental noise becomes white.  In this
limit, decoherence for a single oscillator is characterized by the
dimensionless quantity
\begin{equation}
D\equiv 8{\gamma k_B T\over \hbar\Omega^2}\;,
\end{equation}
where $\Omega$ is the frequency of the Brownian oscillator --- which in
our case is a Fourier mode of the quantum field, so that $\Omega =
ck$.  Since all our field modes decouple, we will first focus on a
single mode, and write $A$ and $A'$ without subscripts to refer to its
amplitude.  (To avoid complex numbers, we will assume that we are
discussing Fourier sine and cosine modes, and rectangular
polarizations.)

The single-mode part of the density matrix propagator, in this weak
coupling, high temperature limit, is
\begin{eqnarray}\label{1prop}
\rho(A,A';t)&=&{\Omega\over2\pi\hbar\sin\Omega t}
\int\!dA_idA'_i\,\Bigl(\rho(A_i,A'_i;0)\nonumber\\
& &\qquad\times\exp\Bigl[{i\Omega\over2\hbar\sin\Omega t}
	[(A^2-A'^2 + A_i^2 - A_i'^2)\cos\Omega t
			- 2(AA_i - A'A_i')]\Bigr]\nonumber\\
& &\qquad\times\;\exp-{\Omega D\over4\hbar\sin^2\Omega t}
	\Bigl[\bigl((A-A')^2+(A_i-A_i')^2\bigr)
	(\Omega t -\sin\Omega t\cos\Omega t)\nonumber\\
& &\qquad\qquad\qquad - 2(A-A')(A_i-A_i')
	(\Omega t\cos\Omega t - \sin\Omega t)\Bigr]\;.
\end{eqnarray}
The mixed state density matrix that evolves from any initial squeezed
state, according to (\ref{1prop}), can be diagonalized explicitly.
We present the results for an arbitrary squeezed state in the Appendix;
here we quote only a particularly relevant special case, namely the
one-parameter family of initial states with $\rho(A,A';0) =
\psi(A,\tau)\psi^*(A',\tau)$ for
\begin{equation}\label{psiAt}
\psi(A,\tau) = Z e^{-{\Omega\over2\hbar}\sigma(\tau)A^2}\;.
\end{equation}
Here $Z$ is a normalization constant, and
\begin{eqnarray}\label{sigma}
|\sigma(\tau)|^2 &=& {2\Omega\tau + \sin2\Omega\tau\over
	2\Omega\tau - \sin2\Omega\tau}\nonumber\\
{\rm Im}\bigl(\sigma(\tau)\bigr) &=& {2\sin^2\Omega\tau\over
	2\Omega\tau - \sin2\Omega\tau}\;.
\end{eqnarray}
The quantity $|\sigma(\tau)|^{-1}$ is the ``squeezing factor'' for
these states.

For a given final time $t$, we will consider the initial state
$\psi(A,\tau)\vert_{\tau=t}$.  By the final time, this state will have
evolved into a state with the density matrix
\begin{eqnarray}\label{rhot}
\rho(A,A';t)&=& \sqrt{\Omega{\rm Re}(\sigma)\over\pi\hbar\Lambda}
\exp-\Bigl({\Omega{\rm Re}(\sigma)\over4\hbar\Lambda}\Bigl[(A+A')^2
	+ \Lambda^2 (A-A')^2\nonumber\\
& &\qquad\qquad\qquad\qquad -2i\bigl(D\sin^2\Omega t
	+ {\rm Im}(\sigma)\bigr)(A^2-A'^2)\Bigr]\Bigr)\nonumber\\
&=& {2\over\Lambda +1}
	e^{i{\Omega{\rm Re}(\sigma)\over2\hbar\Lambda}
	\bigl(D\sin^2\Omega t + {\rm Im}(\sigma)\bigr)(A^2-A'^2)}
	\sum_{n=0}^\infty
	\Bigl({\Lambda-1\over\Lambda+1}\Bigr)^n\phi_n(A)\phi^* n(A')
			\;,
\end{eqnarray}
where $\Lambda \equiv 1 + D\sqrt{(\Omega t)^2 - \sin^2\Omega t}$.  The
$\phi_n$ happen to be the energy eigenfunctions of a harmonic
oscillator with natural frequency $\omega = \Omega {\rm Re}(\sigma)$:
\begin{equation}\label{eigen}
-\hbar^2{d^2\ \over dA^2}\phi_n(A)
	+ [\Omega{\rm Re}(\sigma)]^2 A^2\phi_n(A)
= (2n+1)\hbar\Omega{\rm Re}(\sigma)\phi_n(A)\;.
\end{equation}

This precise form of $\rho(A,A';t)$ has the convenient property that
\begin{eqnarray}\label{larry}
\langle A|\ln\hat{\rho}(t)|A'\rangle &=& \exp\Bigl[i{\Omega{\rm
Re}(\sigma)\over2\hbar\Lambda}\bigl(D\sin^2\Omega t
	+ {\rm Im}(\sigma)\bigr)(A^2-A'^2)\Bigr]\nonumber\\
& &\qquad\qquad\qquad\times
	\Bigl(C_1 - C_2\Bigl[\hbar^2{d^2\ \over dA'^2}
	- \bigl(\Omega{\rm Re}(\sigma)\bigr)^2 A'^2\Bigr]\Bigr)
	\,\delta(A-A')\;,
\end{eqnarray}
where $C_1$ and $C_2$ are constants that may readily be computed from
Eqn. (\ref{rhot}).

We can also determine from Eqn. (\ref{1prop}) the operator valued
functional ${\delta\hat{\rho}(t)\over\delta\psi_i(A_i)}$, for any
$\psi_i$.  Even where the initial state is our special squeezed state
$\psi(A_i,\tau)\vert_{\tau=t}$, carefully chosen with regard to the
final time $t$, this operator variation is somewhat complicated.  Its
diagonal matrix elements, though, are quite simple:
\begin{equation}\label{curly}
\langle A|{\delta\hat{\rho}(t)\over\delta\psi_i(A_i)}|A\rangle
= \sqrt{\Omega\over2\pi\hbar\eta(t)}
	\psi^*(A_i,t)\,e^{-{\Omega\over2\hbar\eta(t)}
	[A-(\cos\Omega t - i\sigma^*(t)\sin\Omega t)A_i]^2}\;,
\end{equation}
where $\eta(t)\equiv \sigma^*\sin^2\Omega t +{D\over2}(\Omega t -
\sin\Omega t\cos\Omega t) - i\sin\Omega t\cos\Omega t$.  It is easy to
see that the property (\ref{curly}) reflects conservation of ${\rm
Tr}\hat{\rho}$.

The variation also has another property, much less trivial (and with
a much more tedious derivation):
\begin{eqnarray}\label{moe}
\lefteqn{\Bigl[\bigl(\hbar^2{d^2\ \over dA^2}
		- [\Omega{\rm Re}(\sigma)]^2 A^2\bigr)
	e^{i{\Omega{\rm Re}(\sigma)\over2\hbar\Lambda}
	\bigl(D\sin^2\Omega t + {\rm Im}(\sigma)\bigr)(A^2-A'^2)}\,
	\langle A|
		{\delta\hat{\rho}(t)\over\delta\psi_i(A_i)}
			|A'\rangle
	\Bigr]_{A=A'}}\nonumber\\
&=& \Bigl(C_3(t)\Bigl[A -A_i(\cos\Omega t
	- i\sigma^*\sin\Omega t)\Bigr]-C_4(t) A\Bigr)\nonumber\\
& &\qquad\qquad\times
	\Bigl[A -A_i(\cos\Omega t - i\sigma^*\sin\Omega t)\Bigr]
	\nonumber\\
&&\qquad\qquad\qquad\qquad\times\,
	\exp-\Bigl[{\Omega\over2\hbar\eta(t)}
	[A-(\cos\Omega t - i\sigma^*(t)\sin\Omega t)A_i]^2\Bigr]
\;,
\end{eqnarray}
where $C_3(t)$ and $C_4(t)$ are functions whose exact form will be
irrelevant to our discussion.

Combining Eqns. (\ref{larry}), (\ref{curly}), and (\ref{moe}),
we find that
\begin{eqnarray}
\lefteqn{\int\!dAdA'\,\langle A'|[1+\ln\hat{\rho}(t)]|A\rangle
\langle A|{\delta\hat{\rho}(t)\over\delta\psi_i(A_i)}|A'\rangle}
\nonumber\\
&=& [1+C_1(t) -{\hbar\eta(t)\over\Omega}C_2(t)C_3(t)]\psi^*(A_i,t)\;.
\end{eqnarray}
This is the constrained Euler-Lagrange equation (\ref{Smin}); the
initial state $\psi(A,t)$ of Eqn. (\ref{psiAt}) therefore minimizes the
entropy of the reduced density matrix at time $t$.  This is identical
to the result obtained in Ref. \cite{ZHP} for the initial state
which minimizes linear entropy at time $t$.

{}From Eqn. (\ref{1prop}), it is apparent that the phase space
translation
\begin{equation}
|\psi_i\rangle \to e^{{i\over\hbar}a\hat{p}_A}
e^{-{i\over\hbar}b\hat{A}} |\psi_i\rangle\;,
\end{equation}
where $\hat{p}_A$ is the canonical momentum operator conjugate to
$\hat{A}$, leads to a unitary transformation of the density operator at
time $t$:
\begin{eqnarray}\label{V}
\hat{\rho}(t) &\to& \hat{V}\hat{\rho}(t)\hat{V}^\dagger\nonumber\\
\hat{V} &=& e^{{i\over\hbar}({p\over\Omega}\sin\Omega t
			+ x\cos\Omega t)\hat{p}_A}
	e^{-{i\over\hbar}(x\cos\Omega t - p\Omega\sin\Omega t)\hat{A}}
	\;.
\end{eqnarray}
The entropy of the state at time $t$ is thus invariant under such phase
space translations of the initial state.  Therefore, the two-parameter
set of initial wave functions $e^{-{i\over\hbar}bA}\psi(A-a,t)$ also
minimize $S(t)$.  We conjecture that these are the only such minimizing
states.

There is thus no initial pure state which will have minimum entropy at
all times.  However, after a few dynamical times, the squeezing and the
imaginary part of $\sigma(t)$ become steadily less significant.  Also,
the state which instantaneously minimizes $S(t)$ oscillates back and
forth, over time, around unsqueezed coherent states.  It is therefore
clear that the least mixing initial states, on average over a few
dynamical times, are the coherent states.  While it is only our special
limit $\gamma\to 0$, $T\to\infty$ that has allowed us to implement the
predictability sieve analytically, calculations in other
models\cite{Gallis}, as well as general arguments\cite{ZHP}, support
the conclusion that coherent states can be considered the natural
pointer states for harmonic oscillators coupled linearly to an
environment.  From the more general analysis of our first section, it
then follows that they are the natural pointer states of an
electromagnetic field mode in a linear medium.

Since the field modes are decoupled, an initial direct product state of
all modes will evolve into a final direct product of mixed states, for
which the total entropy will be the sum of the individual entropies.
It is therefore clear that the generalization of Equation (\ref{Smin})
to all $8(K+1)^3$ decoupled field modes is solved by a direct product
of such squeezed states, and that coherent states of all $8(K+1)^3$
oscillators are the optimum pointer states for the field in a
homogeneous medium.  Furthermore, it follows trivially from Equation
(\ref{V}) that the c-number parameters $x(t),p(t)$ labelling the
pointer states obey the classical equations of motion.  As long as
environmental noise is not so strong that the Gaussian peak in
(\ref{rhot}) becomes too broad too fast, it is clear that classical
mechanics provides a good effective description of the evolution of the
pointer states.  (Of course, the existence of classical histories
follows so trivially from our instantaneous definition of pointer
states only because the dynamics of our model is linear.)

While coherent states of single oscillators are typically interpreted
as localized particles, a coherent state of a quantum field is a vacuum
state displaced by an `external' or `background' field configuration.
The localization associated with decoherence occurs not in the
positions of particles, but in the amplitudes of field modes.  In this
way one can understand that the classical physics which emerges from
quantum electrodynamics, in the presence of a linear environment
environment, will naturally be a field theory and not a many-body
particle theory.

We emphasize that this result is a significant addition to the
observation that one can obtain field equations as classical limits of
quantum dynamics.  After all, the equation of motion for a quantum
harmonic oscillator is exactly the same as it is in the classical case,
but this does nothing towards providing a classical interpetation for a
``Schr\"odinger's Cat'' state.  One must consider decoherence in order
to establish the crucial additional point that the pointer states of
the quantum system, in the presence of an environment, behave in a
sufficiently classical manner.  In the present problem, we have done
this, and observed that the emergent pointer states are classical field
configurations --- a fact which is empirically familiar, but does not
follow at all from the free quantum field theory.

We have therefore made contact, via the predictability sieve, between
decoherence in quantum Brownian motion and the standard field theoretic
notion of a classical background field.  We have confirmed that such
background fields really do behave classically, in that quantum
interference between distinct background field configurations is
rapidly eliminated by a dissipative medium, and that the coherent
quantum states labelled by these background fields are themselves the
states least affected by decoherence.  We now turn to the other
side of the field theoretic coin, and consider how photons excited
above a background field may be affected by the environment.

\section{Quantum halos}

The first point to be made is that our model for the environment is not
intended to describe a sensitive detector.  It is a very poor model for
an ultra-high-gain amplifier, such as is required to detect single
quanta.  So while our discussion concerns the emergence of classical
electrodynamics, we do not really address quantum measurement itself.
Leaving aside the issue of actually detecting photons, however, we
still have a point to address.  Before a photon reaches such a special
environment as a film plate, we know from several classic experiments
that it maintains quantum coherence, despite propagating through air or
other media that are described by our model.  A naive application of
one-particle results to the case of a photon might make this seem
problematic, but in fact the explanation is very simple.

We have found that coherent states are decohered least, on average over
several dynamical times, of all initial pure states.  For a single
harmonic oscillator evolving under (\ref{1prop}), it is well known that
a ``Schr\"odinger's Cat state''
\begin{equation}
|\psi\rangle = c_1 e^{-{i\over\hbar}p_1\hat{A}}
		e^{{i\over\hbar}x_1\hat{p}_A}|0\rangle
	     + c_2 e^{-{i\over\hbar}p_2\hat{A}}
		e^{{i\over\hbar}x_2\hat{p}_A}|0\rangle
\end{equation}
formed by superposing two coherent states, {\it well separated in
$(a,b)$-space}, will decohere thoroughly and rapidly.  At a time
$t={2n\pi\over\Omega}$, the reduced density matrix that has evolved
from this initial state will be
\begin{eqnarray}\label{rho2npi}
\rho(Q,Q';{2n\pi\over\Omega}) &=&
\Bigl[{M\Omega\over\hbar\pi (1+2n\pi D)}\Bigr] \sum_{i,j=1}^2
\exp \Bigl(-{M\Omega\over4\hbar (1+2n\pi D)} {\cal R}_{ij}\Bigr)
	\nonumber\\
{\cal R}_{ij} &=& \Bigl[Q+Q' - x_i - x_j\Bigr]^2
	+ (1+2n\pi D)^2\Bigl[Q-Q' - {x_i-x_j\over1+2n\pi D}\Bigr]^2
	\nonumber\\
& &\qquad -{4i\over M\Omega}\Bigl[Qp_i - Q'p_j
		+ n\pi D (Q-Q')(p_i+p_j)
		- {x_ip_i - x_jp_j\over2}\Bigr]\nonumber\\
& &\qquad + 2n\pi D\Bigl[(x_i-x_j)^2
	+ \bigl({1\over M\Omega}\bigr)^2(p_i-p_j)^2\Bigr]\;.
\end{eqnarray}

{}From the last line of (\ref{rho2npi}), we can infer that the timescale
for decoherence of the two pointer states states is
\begin{equation}\label{tD}
t_D = \Bigl[\Omega D (\Delta^2 -1)\Bigr]^{-1}\;,
\end{equation}
where
\begin{equation}\label{Delta}
\Delta^2 \equiv {M\Omega\over2\hbar}\Bigl[\bigl(x_1-x_2\bigl)^2
+ \bigl({p_1-p_2\over M\Omega}\bigr)^2\Bigr]\;.
\end{equation}
It is obvious that equation (\ref{tD}) makes sense only when
$\Delta^2 >1$.  (We will consider below what happens to a superposition
of two orthogonal states whose wave functions are concentrated within
a phase-space distance of order $\Delta^2 =1$.)

It is also obvious that during processes that occur over timescales
shorter than some $t_{max}$, quantum coherence between two superposed
states will {\it not} decay significantly, if the Wigner functions for
the two states are concentrated within a phase space disc of radius
$\sim 2\sqrt{t_{max}/\Omega D}$.  This rather elementary fact is of
considerable conceptual importance, as it clearly exhibits the
limitations of environmental decoherence.  From it, we can deduce a
succinct refinement of our formulation of environmental-induced
superselection, introducing a new term that complements the notion of a
`pointer state': {\it Every pointer state is surrounded, in Hilbert
space, by a `quantum halo' of states which are not sharply
distinguished from it by the environment.}

The size of the quantum halo of a pointer state is in general a
function both of the strength of environmental noise, and of the
maximum timescale over which it is allowed to act.  However, there is
an upper bound to this timescale, past which the whole notion of
environmentally-induced superselection breaks down anyway, and neither
pointer states nor quantum halos are particularly meaningful.  We can
deduce from Eqn. (\ref{rhot}) that the entropy for an initially
coherent state after $n$ periods of motion is
\begin{equation}\label{S0}
S({n\pi\over\Omega}) = \bigl(1+{n\pi D\over2}\bigr)
		\ln\bigl(1+{n\pi D\over2}\bigr)
		- {n\pi D\over2}\ln\bigl({n\pi D\over2}\bigr)\;.
\end{equation}
When $nD={2\over\pi}$, the entropy even for a pointer state is equal to
that of a statistical mixture of four equally probable pure states.  It
is clear that decoherence this severe does not produce superselection,
but merely swamps the system with environmental noise.  For our pointer
states to be meaningful, therefore, we must have $D\Omega t<<1$.

This means that, as long as decoherence is mild enough to be achieving
superselection instead of mere randomization, the quantum halo
surrounding a coherent state is bound to extend to at least a radius
$\Delta \sim 1$.  Even this minimal halo supports a two dimensional
subspace of states:  the first excited energy eigenstate of the
oscillator resides within the quantum halo of the ground state (see
Figure 1), and a similar halo state may easily be found for any
coherent state.

Generalizing straightforwardly from the single oscillator to the
electromagnetic field in a homogeneous medium, we can conclude that
every background field configuration is surrounded by a quantum halo of
photons.  This explains why a dielectric medium does little to
eliminate quantum interference in a double slit experiment, and why
propagation through an environment will not necessarily destroy the
long-range entanglements of an EPR pair.

\section{Conclusion}

The pointer states of the quantum electromagnetic field, propagating in
a homogeneous linear dielectric medium, are coherent states.  When
decoherence is not so strong that it merely swamps the field with
noise, coherent states evolve almost freely.  The pointer states
therefore behave as classical field configurations, evolving under the
classical equations of motion.  We have therefore provided an insight
into the emergence of classical electromagnetism from quantum
electrodynamics.

Each pointer state of a quantum field is surrounded, in Hilbert space,
by a quantum halo --- a set of states which are negligibly decohered
from the pointer state over whatever time period is of interest.  When
the environmental noise is weak enough that it does not significantly
degrade the pointer states themselves, this quantum halo is large
enough to contain at least a few particles, excited above the background
classical field configuration represented by the pointer state.  We
have thus recovered the familiar field-theoretic dichotomy between
background classical fields and $N$-particle excitations.  The relative
immunity of the particle excitations to decoherence, in comparison
with the strong decoherence of superpositions of distinct pointer
states, explains the co-existence of effective classical
electrodynamics and coherent propagation of photons.

The $n$-particle excitations are not localized by our homogeneous
environment.  All localization occurs in the space of Fourier mode
amplitudes, and not in position space.  This result is consistent with
the ``indications'' arrived at in the studies by K\"ubler and
Zeh\cite{KublerZeh}, and by Kiefer\cite{Kiefer}; but it is in strong
contrast with what one might expect decoherence to do, based on a naive
translation of the particles studied in quantum Brownian motion into
field quanta.  Although a linear dielectric medium does not have the
avalanche instability of a cloud chamber, one would still look to the
concept of decoherence for a general explanation of why electrons, for
example, should behave in the classical limit as localized particles.
Indeed, we do expect that this is the case: the linearly coupled field
we have analyzed in this paper differs in an essential way from the
environmental coupling of a typical matter field.

For matter fields, the interaction Hamiltonian with an environment
tends to be bilinear, rather than linear, in creation and annihilation
operators.  The crude rule of thumb, that pointer states should be
eigenstates of operators that approximately commute with the interaction
Hamiltonian, suggests then that the pointer states for matter fields
should be $n$-particle states rather than coherent states.  And while
a photon can only impart information to a localized environmental
degree of freedom by being absorbed by it, material particles can
scatter, surviving the information transfer without having to rely
on a rare recurrence event to re-emit them.

This observation supplements the usual reference to the statistics of
fermions and bosons, since even a charged scalar field would be
expected to have particle, rather than field, pointer states.  Finally,
we point out that the field-like nature of a Bose condensate of atoms
must be examined with proper consideration for the dynamical origin of
the chemical potential, which can be considered to mimic a linear
interaction (capable of creating or annihilating particles) with an
unobserved environment.

\section{Appendix}

The reduced density matrix $\rho(Q,Q';t)$ which
evolves under the propagator (\ref{1prop}), from an initial squeezed state
\begin{equation}
\langle Q|\psi_I\rangle = \Bigl({M\Omega{\rm
Re}(C)\over\pi\hbar}\Bigr)^{1\over4} e^{-{M\Omega\over2\hbar}C Q^2}
\end{equation}
with complex $C$, is given by
\begin{equation}
\rho(Q,Q';t) = \sqrt{{M\Omega{\rm Re}(C)\over\pi\hbar}}
\sqrt{\alpha(t)} e^{-{M\Omega\over4\hbar}\alpha(t)[(Q+Q')^2
	+ \beta(t)(Q-Q')^2 -2i\lambda(t)(Q^2-Q'^2)]}\;.
\end{equation}
The dimensionless functions $\alpha$, $\beta$, and $\lambda$ are
defined as
\begin{eqnarray}
[\alpha(t)]^{-1} &\equiv& [{\rm Re}(C)]^2\sin^2\Omega t
	+ D{\rm Re}(C)(\Omega t - \sin\Omega t\cos\Omega t) +
	[{\rm Im}(C)\sin\Omega t - \cos\Omega t]^2\;;\nonumber\\
\beta(t) &\equiv& {\rm Re}(C)[1 + D^2(\Omega^2t^2 - \sin^2\Omega t)]
	+ D |C|^2 (\Omega t - \sin\Omega t\cos\Omega t)\nonumber\\
&&\qquad\qquad\qquad	+ D(\Omega t +\sin\Omega t\cos\Omega t)
	- 2D{\rm Im}(C)\sin^2\Omega t\;;\nonumber\\
\lambda(t) &\equiv& [|C|^2 - 1]\sin\Omega t \cos\Omega t
	- {\rm Im}(C)\cos2\Omega t + D{\rm Re}(C)\sin^2\Omega t\;.
\end{eqnarray}

Final states evolving from other initial squeezed states may be
obtained trivially from the result we exhibit, by applying translation
operators as discussed in Section III.

\section{Acknowledgements}

J.R.A. acknowledges the support of the Natural Sciences and Engineering
Research Council of Canada.  Both authors are grateful to Juan Pablo
Paz for valuable discussions.

\vfill\eject

\proclaim Figure Captions.

Figure 1

Plots of the reduced density matrix $\rho(Q,Q';t)$ at $t=0$ (figs. 1a
and 1a') and at $t = 2n\pi\Omega^{-1}$, with $D$ chosen so that $2n\pi
D = 0.05$ (figs. 1b and 1b').  The horizontal axes measure $Q$ and
$Q'$, in units of $\hbar\over M\Omega$.  Figures 1a and 1b show the
evolution of a superposition of two coherent states
$\sqrt{2}\cos(\hat{p}/\sqrt{\hbar M\Omega})|0\rangle$.  It is clear
that decoherence is much advanced in Fig. 1b, but that the two diagonal
peaks are essentially intact, as they represent pointer states.  In
contrast, Figures 1a' and 1b' show the evolution of an initial
superposition of energy eignenstates, ${1\over\sqrt2}(|0\rangle +
|1\rangle)$.  The first excited state lies within the quantum halo of
the ground state, and the superposition has not suffered any
discernible loss of coherence.

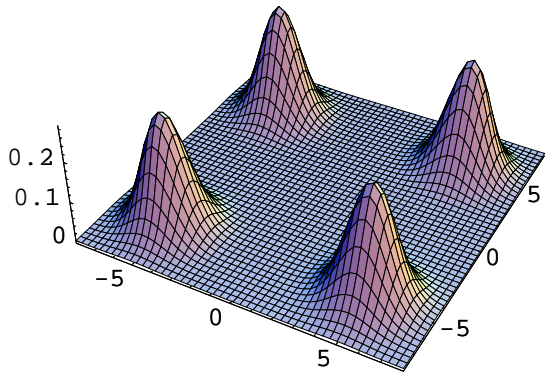
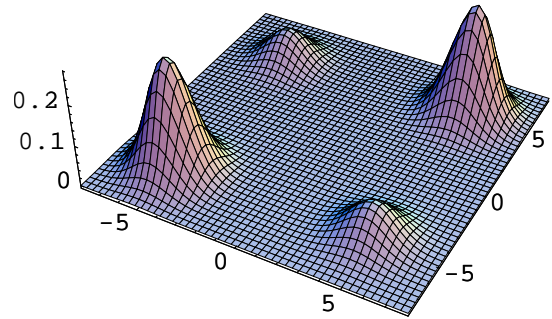
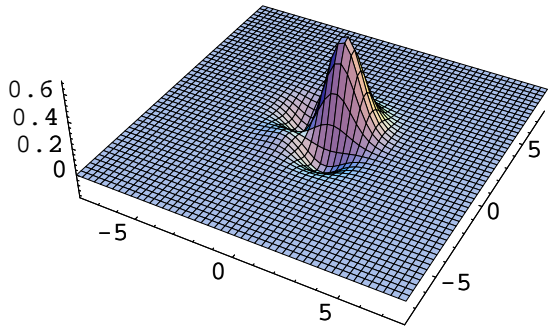
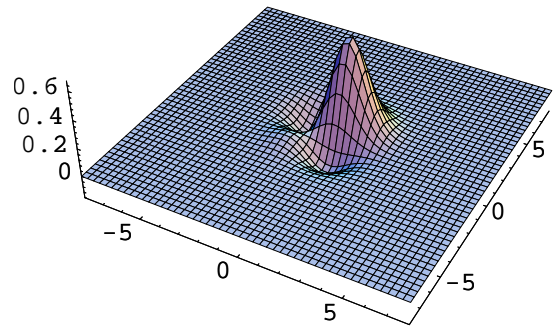
\end{document}